\documentclass[english]{revtex4}
\usepackage[T1]{fontenc}
\usepackage[latin9]{inputenc}
\usepackage{amsmath}
\usepackage{amssymb}

\makeatletter
\@ifundefined{textcolor}{}
{%
 \definecolor{BLACK}{gray}{0}
 \definecolor{WHITE}{gray}{1}
 \definecolor{RED}{rgb}{1,0,0}
 \definecolor{GREEN}{rgb}{0,1,0}
 \definecolor{BLUE}{rgb}{0,0,1}
 \definecolor{CYAN}{cmyk}{1,0,0,0}
 \definecolor{MAGENTA}{cmyk}{0,1,0,0}
 \definecolor{YELLOW}{cmyk}{0,0,1,0}
}

\usepackage{babel}

\makeatother

\usepackage{babel}
\begin{document}
\title{No point-localized photon states}
\author{Scott E. Hoffmann}
\address{School of Mathematics and Physics,~\\
 The University of Queensland,~\\
 Brisbane, QLD 4072~\\
 Australia}
\email{scott.hoffmann@uqconnect.edu.au}

\begin{abstract}
The aim of this paper is to critically examine claims that it is possible
to construct point-localized state vectors for the photon. We supply
a brief proof of the impossibility of this. Then it is found that
the authors making these claims use a non-standard scalar product,
not equal to the quantum-mechanical one. This alternative scalar product
is found to be proportional to a Dirac delta function in position
for the state vectors they use, but the remaining elements of the
proof, namely satisfying all three Newton-Wigner criteria, are lacking.
\end{abstract}
\maketitle

\section{Introduction}

Newton and Wigner \cite{Newton1949} defined three physical criteria
that must be satisfied by the point-localized state vectors of a massive
or massless particle. The first is that a nonzero spatial translation
of a localized state vector must produce another localized state vector
that is orthogonal to the original. Then a rotation of a state vector
localized at the spatial origin at any time, $t,$ must produce another
state vector localized at the origin. Such a state vector can be labelled
by another quantum number that must carry an irreducible representation
of rotations, with state vectors for different values of this quantum
number being mutually orthogonal. Lastly, the scalar product of a
localized state vector and any boost of this state vector must be
a continuous function of the boost velocity.

It is not possible to satisfy all three criteria for the photon. The
reason is not the masslessness of the photon but because of its limited
helicity spectrum, as we will confirm below. This point was noted
by Wightman \cite{Wightman1962}. Within this limitation, it is possible
to define measures of partial (not point) localization for the photon
\cite{Hoffmann2020a}.

Given the impossibility result, it was surprising to learn that some
authors \cite{Hawton2019,Babaei2017,Hawton2017,Mostafazadeh2006}
claim to have constructed point-localized states for the photon. Perhaps
the result was not derived in suffficient detail, so we remedy that
situation in Section \ref{sec:Impossibility-of-point-localized}.
Then, in Section \ref{sec:Analysis-of-claimed}, we closely examine
some of the claims of point-localized photons and attempt to reproduce
their results. Conclusions follow in Section \ref{sec:Conclusions}.

\section{\label{sec:Impossibility-of-point-localized}Impossibility of point-localized
photon state vectors}

As stated above, the helicity spectrum of a massless particle determines
whether it can be point-localized. A hypothetical particle of zero
mass and zero helicity could be localized at a point, with state vectors
\begin{equation}
|\,x,0\,\rangle=\frac{1}{(2\pi)^{\frac{3}{2}}}\int\frac{d^{3}k}{\sqrt{\omega}}\,|\,k,0\,\rangle\,e^{ik\cdot x},\label{eq:1}
\end{equation}
where $\omega=|\boldsymbol{k}|$ and the covariant normalization of
the momentum-helicity eigenvectors (for general helicity, $\lambda$),
\begin{equation}
\langle\,k_{1},\lambda_{1}\,|\,k_{2},\lambda_{2}\,\rangle=\delta_{\lambda_{1}\lambda_{2}}\omega_{1}\delta^{3}(\boldsymbol{k}_{1}-\boldsymbol{k}_{2}),\label{eq:2}
\end{equation}
will be used throughout this paper. Also, in this paper, we use Heaviside-Lorentz
units, in which $\hbar=c=\epsilon_{0}=\mu_{0}=1.$ Only positive energies
are used in this superposition. These state vectors have the equal-time
scalar product
\begin{equation}
\langle\,(t,\boldsymbol{x}_{1}),0\,|\,(t,\boldsymbol{x}_{2}),0\,\rangle=\delta^{3}(\boldsymbol{x}_{1}-\boldsymbol{x}_{2})\label{eq:3}
\end{equation}
and rotate and translate as expected for localized state vectors.
They boost continuously.

The photon has only two physical helicities, $\lambda=\pm1$. If there
were a $\lambda=0$ photon in addition to these two, we could construct
three state vectors (for $\sigma=-1,0,1$)
\begin{equation}
|\,x,\sigma,3\,\rangle=\frac{1}{(2\pi)^{\frac{3}{2}}}\int\frac{d^{3}k}{\sqrt{\omega}}\sum_{\lambda=-1,0,1}|\,k,\lambda\,\rangle\,\mathcal{R}_{\lambda\sigma}^{(1)-1}[\hat{\boldsymbol{k}}]\,e^{ik\cdot x}.\label{eq:4}
\end{equation}
Here $\mathcal{R}_{\lambda\mu}^{(1)-1}(\hat{\boldsymbol{k}})$ are
$j=1$ matrix elements of the inverse of the standard rotation
\begin{equation}
R_{0}[\hat{\boldsymbol{k}}]=R_{z}(\varphi)R_{y}(\theta)R_{z}(-\varphi)\label{eq:5}
\end{equation}
that takes $\hat{\boldsymbol{z}}$ into the momentum direction $\hat{\boldsymbol{k}}=(\theta,\varphi).$

The basis vectors rotate as
\begin{equation}
U(R)\,|\,k,\lambda\rangle=|\,R\,k,\lambda\,\rangle\,e^{-i\lambda w(R,\hat{\boldsymbol{k}})},\label{eq:6}
\end{equation}
with a Wigner rotation, a rotation about the momentum direction, $\hat{\boldsymbol{k}},$
specified by \cite{Hoffmann2020a} 
\begin{equation}
R(w(R,\hat{\boldsymbol{k}})\,\hat{\boldsymbol{k}})=R_{0}^{-1}[R\hat{\boldsymbol{k}}]\,R\,R_{0}[\hat{\boldsymbol{k}}].\label{eq:7}
\end{equation}
Then we have (no sum over $\lambda$)
\begin{equation}
U(R)\,|\,k,\lambda\,\rangle\,\mathcal{R}_{\lambda\sigma}^{(1)-1}(\hat{\boldsymbol{k}})=\sum_{\sigma^{\prime}=-1}^{1}|\,R\,k,\lambda\,\rangle\,\mathcal{R}_{\lambda\sigma^{\prime}}^{(1)-1}[R\hat{\boldsymbol{k}}]\,\mathcal{D}_{\sigma^{\prime}\sigma}^{(1)}(R).\label{eq:8}
\end{equation}
This is an important result, as it shows that each helicity rotates
with the same transformation matrix.

This leads to the required rotation behaviour
\begin{equation}
U(R)\,|\,(t,\boldsymbol{x}),\sigma,3\,\rangle=\sum_{\mu^{\prime}=-1}^{1}|\,(t,R\,\boldsymbol{x}),\sigma^{\prime},3\,\rangle\,\mathcal{D}_{\sigma^{\prime}\sigma}^{(1)}(R),\label{eq:9}
\end{equation}
according to the unitary irreducible rotation representation with
angular momentum quantum number $j=1.$ It is clear that this construction
(Eq. (\ref{eq:4})) is the unique solution (up to phase changes dependent
on helicity) of the requirement in Eq. (\ref{eq:9}).

The unitary transformation
\begin{equation}
\langle\,\sigma\,|\,i\,\rangle=\sqrt{\frac{4\pi}{3}}\,Y_{1\sigma}^{*}(\hat{\boldsymbol{i}})\quad\mathrm{for}\ i=1,2,3,\label{eq:10}
\end{equation}
produces three state vectors for every $x,$
\begin{equation}
|\,x,i,3\,\rangle=\sum_{\sigma=-1}^{1}|\,x,\sigma,3\,\rangle\,\langle\,\sigma\,|\,i\,\rangle\quad\mathrm{for}\ i=1,2,3.\label{eq:11}
\end{equation}
We note that the coefficients in the superposition can be identified
as the components of complex conjugate polarization vectors in a particular
gauge,
\begin{equation}
\epsilon_{i}^{*}(\hat{\boldsymbol{k}},\lambda)=\sum_{\sigma=-1}^{1}\mathcal{R}_{\lambda\sigma}^{(1)-1}[\hat{\boldsymbol{k}}]\,\langle\,\sigma\,|\,i\,\rangle=\sum_{\sigma=-1}^{1}\sum_{j=1}^{3}\sum_{k=1}^{3}\langle\,\lambda\,|\,j\,\rangle\,R_{0jk}^{-1}[\hat{\boldsymbol{k}}]\,\langle\,k\,|\,\sigma\,\rangle\langle\,\sigma\,|\,i\,\rangle=\sum_{j=1}^{3}R_{0ij}[\hat{\boldsymbol{k}}]\,\langle\,\lambda\,|\,j\,\rangle.\label{eq:12}
\end{equation}
They are obtained by a rotation from their $\hat{\boldsymbol{k}}=\hat{\boldsymbol{z}}$
values
\begin{equation}
\boldsymbol{\epsilon}^{*}(\hat{\boldsymbol{z}},+1)=-\frac{1}{\sqrt{2}}\begin{pmatrix}1\\
-i\\
0
\end{pmatrix},\quad\epsilon^{*}(\hat{\boldsymbol{z}},0)=\begin{pmatrix}0\\
0\\
1
\end{pmatrix},\quad\epsilon^{*}(\hat{\boldsymbol{z}},-1)=+\frac{1}{\sqrt{2}}\begin{pmatrix}1\\
+i\\
0
\end{pmatrix}.\label{eq:13}
\end{equation}
So we can write
\begin{equation}
|\,x,i,3\,\rangle=\frac{1}{(2\pi)^{\frac{3}{2}}}\int\frac{d^{3}k}{\sqrt{\omega}}\sum_{\lambda=-1,0,1}|\,k,\lambda\,\rangle\,\epsilon_{i}^{*}(\hat{\boldsymbol{k}},\lambda)\,e^{ik\cdot x}.\label{eq:14}
\end{equation}

The three state vectors in Eq. (\ref{eq:14}) rotate according to
the three-vector representation of rotations
\begin{equation}
U(R)\,|\,(t,\boldsymbol{x}),i,3\,\rangle=\sum_{i^{\prime}=-1}^{1}|\,(t,R\,\boldsymbol{x}),i^{\prime},3\,\rangle\sum_{\sigma^{\prime},\sigma=-1}^{1}\langle\,i^{\prime}\,|\,\sigma^{\prime}\,\rangle\mathcal{D}_{\sigma^{\prime}\sigma}^{(1)}(R)\,\langle\,\sigma\,|\,i\,\rangle=\sum_{i^{\prime}=-1}^{1}|\,(t,R\,\boldsymbol{x}),i^{\prime},3\,\rangle\,R_{i^{\prime}i}.\label{eq:15}
\end{equation}

Both sets of state vectors translate correctly and satisfy the equal-time
orthonormality relations
\begin{equation}
\langle\,(t,\boldsymbol{x}_{1}),\sigma_{1},3\,|\,(t,\boldsymbol{x}_{2}),\sigma_{2},3\,\rangle=\frac{1}{(2\pi)^{3}}\int d^{3}k\,\sum_{\lambda=-1,0,1}\mathcal{R}_{\sigma_{1}\lambda}^{(1)}[\hat{\boldsymbol{k}}]\mathcal{R}_{\lambda\sigma_{2}}^{(1)-1}(\hat{\boldsymbol{k}})\,e^{i\boldsymbol{k}\cdot(\boldsymbol{x}_{1}-\boldsymbol{x}_{2})}=\delta_{\sigma_{1}\sigma_{2}}\delta^{3}(\boldsymbol{x}_{1}-\boldsymbol{x}_{2})\label{eq:16}
\end{equation}
and
\begin{equation}
\langle\,(t,\boldsymbol{x}_{1}),i_{1},3\,|\,(t,\boldsymbol{x}_{2}),i_{2},3\,\rangle=\delta_{i_{1}i_{2}}\delta^{3}(\boldsymbol{x}_{1}-\boldsymbol{x}_{2}),\label{eq:17}
\end{equation}
as required for sets of three point-localized state vectors.

Now, for the physical photon, we must remove the zero helicity in
the new definition
\begin{equation}
|\,x,\sigma,\gamma\,\rangle=\frac{1}{(2\pi)^{\frac{3}{2}}}\int\frac{d^{3}k}{\sqrt{\omega}}\sum_{\lambda=-1,1}|\,k,\lambda\,\rangle\,\mathcal{R}_{\lambda\sigma}^{(1)-1}(\hat{\boldsymbol{k}})\,e^{ik\cdot x}\quad\mathrm{for}\ \sigma=-1,0,1.\label{eq:18}
\end{equation}
We continue to use the $j=1$ rotation representation, as it is expected
to come closest to the desired result. From the result of Eq. (\ref{eq:8}),
we see that these three still rotate in the desired way. However,
the equal-time overlap becomes
\begin{equation}
\langle\,(t,\boldsymbol{x}_{1}),\sigma_{1},\gamma\,|\,(t,\boldsymbol{x}_{2}),\sigma_{2},\gamma\,\rangle=\frac{1}{(2\pi)^{3}}\int d^{3}k\,\sum_{\lambda=-1,1}\mathcal{R}_{\sigma_{1}\lambda}^{(1)}[\hat{\boldsymbol{k}}]\mathcal{R}_{\lambda\sigma_{2}}^{(1)-1}(\hat{\boldsymbol{k}})\,e^{i\boldsymbol{k}\cdot(\boldsymbol{x}_{1}-\boldsymbol{x}_{2})}.\label{eq:19}
\end{equation}
The sum over helicities is
\begin{align}
\sum_{\lambda=-1,1}\mathcal{R}_{\sigma_{1}\lambda}^{(1)}[\hat{\boldsymbol{k}}]\mathcal{R}_{\lambda\sigma_{2}}^{(1)-1}(\hat{\boldsymbol{k}}) & =\delta_{\sigma_{1}\sigma_{2}}-\mathcal{R}_{\sigma_{1}0}^{(1)}[\hat{\boldsymbol{k}}]\mathcal{R}_{0\sigma_{2}}^{(1)-1}(\hat{\boldsymbol{k}})\nonumber \\
 & =\delta_{\sigma_{1}\sigma_{2}}-\begin{pmatrix}\frac{1}{2}\sin^{2}\theta & -\frac{e^{-i\varphi}}{\sqrt{2}}\sin\theta\cos\theta & -\frac{1}{2}e^{-i2\varphi}\sin^{2}\theta\\
-\frac{e^{+i\varphi}}{\sqrt{2}}\sin\theta\cos\theta & \cos^{2}\theta & +\frac{e^{-i\varphi}}{\sqrt{2}}\sin\theta\cos\theta\\
-\frac{1}{2}e^{+i2\varphi}\sin^{2}\theta & +\frac{e^{+i\varphi}}{\sqrt{2}}\sin\theta\cos\theta & \frac{1}{2}\sin^{2}\theta
\end{pmatrix}_{\sigma_{1}\sigma_{2}}.\label{eq:20}
\end{align}
So
\begin{equation}
\langle\,(t,\boldsymbol{x}_{1}),\sigma_{1},\gamma\,|\,(t,\boldsymbol{x}_{2}),\sigma_{2},\gamma\,\rangle=\delta_{\sigma_{1}\sigma_{2}}\delta^{3}(\boldsymbol{x}_{1}-\boldsymbol{x}_{2})-\frac{1}{(2\pi)^{3}}\int d^{3}k\,M_{\sigma_{1}\sigma_{2}}(\hat{\boldsymbol{k}})\,e^{i\boldsymbol{k}\cdot(\boldsymbol{x}_{1}-\boldsymbol{x}_{2})},\label{eq:21}
\end{equation}
where $M_{\sigma_{1}\sigma_{2}}(\hat{\boldsymbol{k}})$ is the matrix
appearing in the second term of Eq. (\ref{eq:20}).

For the three-vector states
\begin{align}
\langle\,(t,\boldsymbol{x}_{1}),i_{1},\gamma\,|\,(t,\boldsymbol{x}_{2}),i_{2},\gamma\,\rangle & =\frac{1}{(2\pi)^{3}}\int d^{3}k\,(\delta_{i_{1}i_{2}}-\sum_{\lambda=-1,1}\langle\,i_{1}\,|\,\sigma_{1}\,\rangle\mathcal{R}_{\sigma_{1}0}^{(1)}[\hat{\boldsymbol{k}}]\mathcal{R}_{0\sigma_{2}}^{(1)-1}(\hat{\boldsymbol{k}})\langle\,\sigma_{2}\,|\,i_{2}\,\rangle)\,e^{i\boldsymbol{k}\cdot(\boldsymbol{x}_{1}-\boldsymbol{x}_{2})}.\label{eq:22}
\end{align}
with summation over $\mu_{1}$ and $\mu_{2}$ implied. We find the
result
\begin{equation}
\langle\,(t,\boldsymbol{x}_{1}),i_{1},\gamma\,|\,(t,\boldsymbol{x}_{2}),i_{2},\gamma\,\rangle=\delta_{i_{1}i_{2}}\delta^{3}(\boldsymbol{x}_{1}-\boldsymbol{x}_{2})-\frac{1}{(2\pi)^{3}}\int d^{3}k\,\hat{\boldsymbol{i}}_{1}\cdot\hat{\boldsymbol{k}}\hat{\boldsymbol{k}}\cdot\hat{\boldsymbol{i}}_{2}\,e^{i\boldsymbol{k}\cdot(\boldsymbol{x}_{1}-\boldsymbol{x}_{2})},\label{eq:23}
\end{equation}
illustrating the impossibility of point-localized photon state vectors,
at least using the $j=1$ rotation representation.

Clearly using $j=0$ gives the wrong rotation properties for the state
vectors. Using $j\geq2,$ there will be unwanted terms inside the
momentum integral for the scalar products, given by
\begin{equation}
T_{\sigma_{1}\sigma_{2}}^{(j)}(\hat{\boldsymbol{k}})=-\sum_{\{\lambda\}}\mathcal{R}_{\sigma_{1}0}^{(j)}[\hat{\boldsymbol{k}}]\mathcal{R}_{0\sigma_{2}}^{(j)-1}(\hat{\boldsymbol{k}}),\label{eq:24}
\end{equation}
a sum over all helicities other than $\lambda=\pm1.$

\section{\label{sec:Analysis-of-claimed}Analysis of claimed point-localized
states}

Hawton \cite{Hawton2019} (her Eq. (34), summed over $\epsilon$
(the sign of the energy) and $\lambda$) constructs state vectors
(using our normalization conventions)
\begin{equation}
|\,A^{\mu}(x)\,\rangle=\frac{1}{(2\pi)^{\frac{3}{2}}}\,\int\frac{d^{3}k}{\omega}\sum_{\lambda=\pm1}|\,k,\lambda\,\rangle\,\epsilon^{*\mu}(\hat{\boldsymbol{k}},\lambda)\,e^{+ik\cdot x}.\label{eq:25}
\end{equation}
Here $\epsilon^{*\mu}(\hat{\boldsymbol{k}},\lambda)$ is now a four-component
polarization vector, with $\mu=0,1,2,3.$

Negative frequency contributions, $|\,A^{(-)\mu}(x)\,\rangle,$ with
the phase factor $\exp(-ik\cdot x),$ are also included, but they
don't translate correctly:
\begin{equation}
U(T(a))\,|\,A^{(-)\mu}(x)\,\rangle=|\,A^{(-)\mu}(x-a)\,\rangle\label{eq:26}
\end{equation}
instead of
\begin{equation}
U(T(a))\,|\,A^{(-)\mu}(x)\,\rangle=|\,A^{(-)\mu}(x+a)\,\rangle.\label{eq:27}
\end{equation}
It seems that these negative frequencies are not essential to their
result, so we omit them here.

The first issue is gauge variation. The four-component polarization
vectors appear in the electromagnetic field strength operators, $\hat{F}^{\mu\nu}(x),$
and in the four-component gauge field operator, $\hat{A}^{\mu}(x).$
A gauge transformation,
\begin{equation}
\epsilon^{\mu}(\hat{\boldsymbol{k}},\lambda)\rightarrow\epsilon^{\mu}(\hat{\boldsymbol{k}},\lambda)+g(|\boldsymbol{k}|)\,k^{\mu},\label{eq:28}
\end{equation}
for arbitrary $g,$ leaves the observable electromagnetic field strengths
unchanged, but changes the gauge field components. We note that such
a gauge transformation preserves the Lorentz condition,
\begin{equation}
k\cdot\epsilon(\hat{\boldsymbol{k}},\lambda)\equiv0,\label{eq:29}
\end{equation}
which is essential for the gauge fields to satisfy the Maxwell equations.

Unless a particular gauge is chosen, Eq. (\ref{eq:25}) does not give
a unique definition. We choose to consider the radiation gauge, with
polarization vectors $\epsilon_{R}^{\mu}(\hat{\boldsymbol{k}},\lambda),$
in which $\epsilon_{R}^{0}(\hat{\boldsymbol{k}},\lambda)\equiv0$
and the spatial parts are given by Eqs. (\ref{eq:12},\ref{eq:13}).
This is part of the Lorentz family of gauges since $\boldsymbol{k}\cdot\boldsymbol{\epsilon}_{R}(\hat{\boldsymbol{k}},\lambda)=\hat{\boldsymbol{z}}\cdot\boldsymbol{\epsilon}_{R}(\hat{\boldsymbol{z}},\lambda)\equiv0.$

Then we arrive at three state vectors for every $x,$
\begin{equation}
|\,x,i,\mathrm{H}\,\rangle=\frac{1}{(2\pi)^{\frac{3}{2}}}\,\int\frac{d^{3}k}{\omega}\sum_{\lambda=\pm1}|\,k,\lambda\,\rangle\,\epsilon_{R}^{*i}(\hat{\boldsymbol{k}},\lambda)\,e^{+ik\cdot x}\quad\mathrm{for}\ i=1,2,3.\label{eq:30}
\end{equation}
Since a different factor of the rotational invariant, $\omega,$ in
the superposition will not change the rotation properties, we see
that these state vectors rotate according to the irreducible representation.
Their equal-time quantum-mechanical scalar products are
\begin{multline*}
\langle\,(t,\boldsymbol{x}_{1}),i_{1},\mathrm{H}\,|\,(t,\boldsymbol{x}_{2}),i_{2},\mathrm{H}\,\rangle=\frac{1}{(2\pi)^{3}}\int\frac{d^{3}k}{\omega}\sum_{\lambda=\pm1}\epsilon_{R}^{i_{1}}(\hat{\boldsymbol{k}},\lambda)\epsilon_{R}^{*i_{2}}(\hat{\boldsymbol{k}},\lambda)\,e^{i\boldsymbol{k}\cdot(\boldsymbol{x}_{1}-\boldsymbol{x}_{2})}\\
=\frac{1}{(2\pi)^{3}}\int\frac{d^{3}k}{\omega}(\delta_{i_{1}i_{2}}-\hat{\boldsymbol{i}}_{1}\cdot\hat{\boldsymbol{k}}\hat{\boldsymbol{k}}\cdot\hat{\boldsymbol{i}}_{2})\,e^{i\boldsymbol{k}\cdot(\boldsymbol{x}_{1}-\boldsymbol{x}_{2})},
\end{multline*}
clearly not giving the required orthogonality result.

It is remarkable that these authors \cite{Hawton2019,Babaei2017,Hawton2017,Mostafazadeh2006}
define an \textit{alternative} scalar product, not equal to the quantum-mechanical
scalar product. If a state vector is defined as a superposition of
basis vectors with known scalar products, then the quantum-mechanical
scalar product of two such state vectors is completely defined and
not subject to arbitrary redefinition. The orthogonality properties
of the basis vectors, in this case, follow from the fact that they
are eigenvectors of Hermitian observables. State vectors can be defined
in different ways, but then their scalar products are fixed.

The alternative scalar product is defined by
\begin{equation}
(\,A_{1}(t,\boldsymbol{x}_{1}),A_{2}(t,\boldsymbol{x}_{2})\,)=\frac{1}{(2\pi)^{3}}\int\frac{d^{3}k}{\omega}\sum_{\lambda=\pm1}\sum_{i=1}^{3}\epsilon_{R}^{i}(\hat{\boldsymbol{k}},\lambda)\epsilon_{R}^{*i}(\hat{\boldsymbol{k}},\lambda)\,e^{i\boldsymbol{k}\cdot(\boldsymbol{x}_{1}-\boldsymbol{x}_{2})}.\label{eq:32}
\end{equation}
To verify the rotation criterion of Newton and Wigner, nine scalar
products for each $t,x_{1},x_{2}$ are needed, one for each pair $i_{1},i_{2}$.
But only one is given here, with the index $i$ summed over. If there
were no sum over $i,$ we would again have
\begin{equation}
\sum_{\lambda=\pm1}\epsilon_{R}^{i_{1}}(\hat{\boldsymbol{k}},\lambda)\epsilon_{R}^{*i_{2}}(\hat{\boldsymbol{k}},\lambda)=\delta_{i_{1}i_{2}}-\hat{\boldsymbol{i}}_{1}\cdot\hat{\boldsymbol{k}}\hat{\boldsymbol{k}}\cdot\hat{\boldsymbol{i}}_{2}.\label{eq:32.1}
\end{equation}

Using
\begin{equation}
\boldsymbol{\epsilon}_{R}(\hat{\boldsymbol{k}},\lambda)\cdot\boldsymbol{\epsilon}_{R}^{*}(\hat{\boldsymbol{k}},\lambda)=\boldsymbol{\epsilon}_{R}(\hat{\boldsymbol{z}},\lambda)\cdot\boldsymbol{\epsilon}_{R}^{*}(\hat{\boldsymbol{z}},\lambda)=1\label{eq:33}
\end{equation}
gives their result
\begin{equation}
(\,A_{1}(t,\boldsymbol{x}_{1}),A_{2}(t,\boldsymbol{x}_{2})\,)=2\,\delta^{3}(\boldsymbol{x}_{1}-\boldsymbol{x}_{2}).\label{eq:34}
\end{equation}

This result does not, as claimed, prove that localized photon state
vectors have been constructed.

\section{\label{sec:Conclusions}Conclusions}

We have confirmed that no point-localized state vectors can be constructed
for the photon that satisfy all three of the criteria of Newton and
Wigner \cite{Newton1949}. The reason is the limited helicity spectrum
of the photon. If there were a zero helicity state in addition to
$\lambda=\pm1,$ such localized states, three for every position and
time, could be constructed.

We considered the claims of other authors \cite{Hawton2019,Babaei2017,Hawton2017,Mostafazadeh2006}
that they could construct point-localized photon states satisfying
the three criteria of Newton and Wigner. It was found that they employ
an alternative definition of scalar product, not equal to the well-defined
quantum-mechanical one. It is only by using this alternative scalar
product that they are able to find a Dirac delta function in position.
But the argument is incomplete. The rotation of their state vectors
involves a subspace of dimension 3. Orthogonality of state vectors
with different values of the index would be required to complete the
proof, but no such result is given and would clearly not be possible.

We note that these authors also claim to have constructed a positive-definite
position probability density for the photon (the zero component of
a locally conserved four-current). If this were the case, it would
have to take the form
\[
\hat{\rho}(x)=\sum_{\sigma}|\,x,\sigma\,\rangle\langle\,x,\sigma\,|.
\]
The basis vectors would have to rotate according to an irreducible
representation and satisfy mutual orthogonality between different
values of the index, $\sigma$. Since we know this is not possible,
the claim is baseless.

\bibliographystyle{vancouver}

\end{document}